# Relativistic angle-changes and frequency-changes

**Eric Baird**  (eric_baird@compuserve.com)

We generate a set of "relativistic" predictions for the relationship between viewing angle and apparent frequency, for each of three different non-transverse shift equations. We find that a detector aimed transversely (in the lab frame) at a moving object should report a redshift effect with two out of the three equations.

## 1. Introduction

One of the advantages of Einstein's special theory [1] over other contemporary models was that it provided clear and comparatively straightforward predictions for angle-changes and frequency changes at different angles. These calculations could be rather confusing in the variety of aether models that were available at the time (*see* e.g. Lodge, 1894 [2] ).

In this paper, we derive the relativistic angle-changes and wavelength-changes that would be associated with each of the three main non-transverse Doppler equations, if used as the basis of a relativistic model.

## 2. Geometry

### Wavefront shape

If a "stationary" observer emits a unidirectional pulse of light and watches the progress of the spreading wavefront, the signal that they detect is not the outgoing wavefront itself, but a secondary (incoming) signal generated when the outgoing wavefront illuminates dust or other objects in the surrounding region.

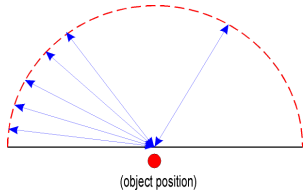
(object position)

If we assumed that the speed of light was constant in the observer's own frame, any dust-particles that the observer saw to be illuminated at the same moment would be said to lie on a spherical surface, and these "illumination-events" would be said to be simultaneous.

In a different inertial frame, the observer's spatial coordinates change while these light-signals are in flight. The description of the same light-rays and illuminated surfaces now involves rays emanating from one spatial position, striking dust-particles, and generating secondary signals which then converge on a different spatial position.

In the new frame, our original illuminated spherical surface now has to obey the condition that the round-trip flight-time from **A**→surface→**B** is the same for rays sent in any direction, and a map of a cross-section through the surface now gives us an elongated ellipse, with the emission and absorption points being the two ellipse foci.

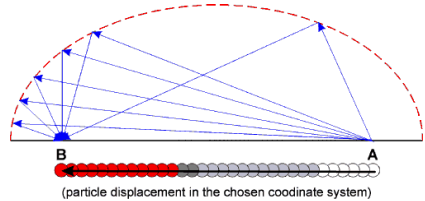
(particle displacement in the chosen coordinate system)

The same set of illumination-events can mark out a spherical spatial surface in one frame and a spheroidal spatial surface in the other. [3][4]

### Ellipse dimensions

Instead of using the "fixed flat aether approach" for calculating distances in our diagram (setting the distance between foci as $v$ and the round-trip distance as $2c$), we will instead construct our map around the wavelength-distances that must be fitted into it, regardless of whether or not these distances are expected to fit nicely into flat spacetime.

The three Doppler formulae used are:

$$freq' / freq = (c-v) / c \qquad \ldots (1)$$

$$freq' / freq = \sqrt{(c-v)/(c+v)} \qquad \ldots (2)$$

$$freq' / freq = c / (c+v) \qquad \ldots (3)$$

, where $v$ is recession velocity

Equation (1) is often associated with emitter theories, (3) is usually associated with an absolute aether stationary in the observer's frame, and (2) is the intermediate prediction used by special relativity (and is the root product of the other two equations). [5]

## 3. *Resulting wavefront maps*

### Ellipse proportions

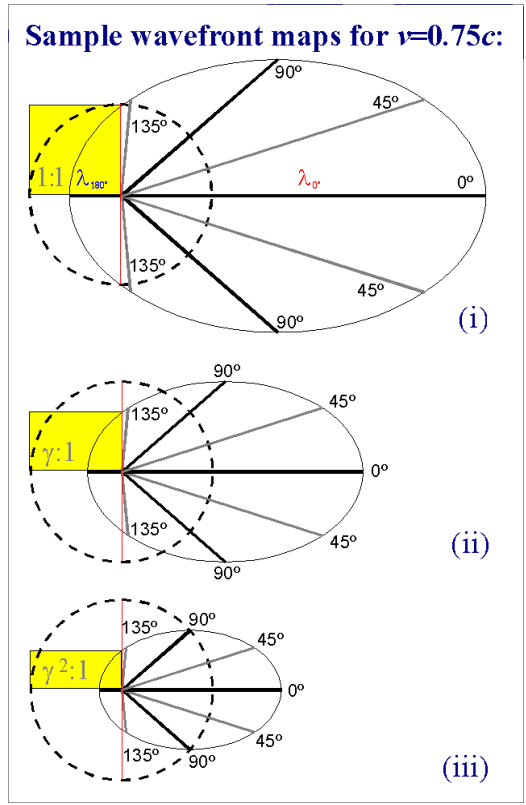

Each of our three Doppler equations (1), (2), (3) has the same ratio $\lambda_{0°}:\lambda_{180°}$, so all three ellipses share the same proportions and angles. Each ellipse is related to its immediate neighbours by a Lorentz magnification or reduction.

### Reading information from the wavefront map

Each wavefront map is drawn for a specific velocity, and includes the reference-angles and wavelengths for a "stationary" object (dashed circular outline) and the corresponding angles and wavelength distances for the moving object (solid ellipse).

Each ray drawn on our map then has two nominal angles (defined by where the ray hits the perimeter of each wavefront outline), and two nominal wavelengths (defined by the distance from the common focus to each wavefront outline).

Thus, for special relativity at $v=0.75c$ (second diagram), we can immediately see that a light-ray aimed at the moving particle at 90° in the background frame will strike the front of the particle at slightly less than 45° from the particle's path, and will have a wavelength that is about two thirds of the original wavelength. Conversely, when the particle emits a ray that is *received* at 90° in the background frame, we can see that the wavelength is magnified about 1½times, and in the particle's frame was aimed rearwards at roughly 45° from the perpendicular.

Precise calculations are given in section 5.

## 4. *Special relativity*

### Lorentz contraction

The elliptical outline of the second map can be converted back into a circle of radius *c* by applying a Lorentz contraction in the direction of motion.

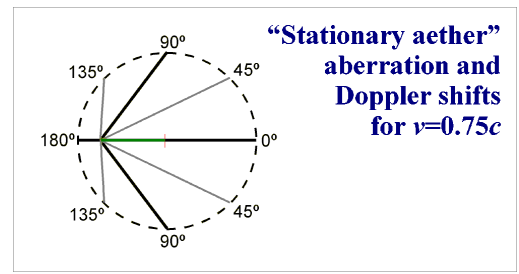

The resulting diagram (above) then shows the *non*-relativistic Nineteenth-Century predictions for angle-changes and wavelength changes associated with an object moving through a flat stationary aether, viewed by a stationary observer (Lodge [2], fig 4, pp.739).

### Minkowski diagrams

We can also obtain special relativity's elliptical wavefront diagram by taking a standard Minkowski light-cone diagram, [6] and slicing it at an angle parallel with the plane of simultaneity of an object in a different frame.

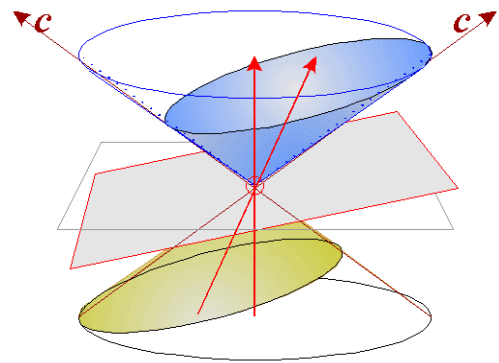

Lorentz transforms can then be used to turn the "skewed" lightcone diagram back into the usual "unskewed" version.

## 5. Resulting formulae

### Angle changes

All three diagrams produce the same angle-change relationships presented in Einstein's 1905 paper [1] as "the law of aberration in its most general form",

$$\cos A' = -\frac{\cos A - v/c}{1 - \cos A (v/c)}$$

### Wavelength changes at all angles

The wavelength-changes in rays arriving in the laboratory frame, if angle A is measured in the lab frame (and $v$ is positive for recession when A=0°), are:

$$\lambda'/\lambda = \frac{1 + (v/c)\cos A}{1 - v^2/c^2} \qquad \text{... (i)}$$

$$\lambda'/\lambda = \frac{1 + (v/c)\cos A}{\sqrt{1 - v^2/c^2}} \qquad \text{... (ii)}$$

$$\lambda'/\lambda = 1 + (v/c)\cos A \qquad \text{... (iii)}$$

The wavelengths generated by the second diagram are those given in section 7 of the 1905 electrodynamics paper. Wavelengths generated by (1) and (3) are simply special relativity's values, multiplied or divided by a Lorentz term.

## 6. "Transverse" shift predictions

### Emitter ® lab

If we aim a detector transversely at a moving object (with our "ninety degrees" being measured in the detector's frame), the predicted wavelength changes seen by the detector are:

$$freq'/freq = 1 - v^2/c^2 \qquad \text{... (i)}$$

$$freq'/freq = \sqrt{1 - v^2/c^2} \qquad \text{... (ii)}$$

$$freq'/freq = 1 \qquad \text{... (iii)}$$

As with the non-transverse predictions (1), (2), (3), special relativity's "transverse" predictions for observed frequencies and apparent ruler-changes are the root product average of the "non-relativistic" predictions made by assuming a fixed speed of light in each frame.

### Lab ® emitter

If a signal is aimed at the moving object at 90° in the lab frame, the wavelength changes seen by the object should be:

$$freq'/freq = 1 \qquad \text{... (i)}$$

$$freq'/freq = 1/\sqrt{1 - v^2/c^2} \qquad \text{... (ii)}$$

$$freq'/freq = 1/(1 - v^2/c^2) \qquad \text{... (iii)}$$

These predictions differ from the earlier set because of our asymmetrical decision to measure all angles in the laboratory frame.

## 7. Round-trip shift predictions

### ... transverse

If we aim a signal at a moving transponder and catch the retransmitted signal, then if both signals paths are transverse to the transponder path in the laboratory frame, multiplying the two previous sets of results together gives us final round-trip predictions of:

$$freq'/freq = 1 - v^2/c^2 \qquad \text{... (i)}$$

$$freq'/freq = 1 \qquad \text{... (ii)}$$

$$freq'/freq = 1/(1 - v^2/c^2) \qquad \text{... (iii)}$$

Equation (1) predicts a Lorentz-squared redshift after two lab-transverse frame transitions, and special relativity gives a null result.

### ... non-transverse

We could also aim our source and detector at the transponder *non*-transversely. Multiplying the approach and recession shifts gives:

$$\frac{freq'}{freq} = \frac{c-v}{c} \times \frac{c-(-v)}{c} = 1 - v^2/c^2 \qquad \text{... (i)}$$

$$\frac{freq'}{freq} = \sqrt{\frac{c-v}{c+v}} \times \sqrt{\frac{c-(-v)}{c+(-v)}} = 1 \qquad \text{... (ii)}$$

$$\frac{freq'}{freq} = \frac{c}{c+v} \times \frac{c}{c+(-v)} = 1/(1 - v^2/c^2) \qquad \text{... (iii)}$$

These round-trip frequency shifts are the same as in the previous "transverse" calculation.

### … at any angle

If we mount our detector and signal source on a turntable, the angle that the source→detector line makes with the transponder path has no effect on the final calculated round-trip shift.

If we aim two lasers collinearly into an ion beam, and measure the non-transverse frequency shifts by measuring how far the lasers need to be detuned in order to achieve resonance in the beam, [7] the relationship between the product of the shifted frequencies and the original frequency should be unaffected by any misalignment of the lasers or spread of observation angles, [8] provided that the angular error is the same for both observations.

## 8. Consequences

### Generality of the E=mc² result

In a previous paper [9] we showed that the E=mc² mass-energy relationship can be derived from the frame-dependent total momentum of two plane waves emitted directly forwards and backwards along the direction of motion, if the non-transverse shift law is (1), with the equivalent calculation using (2) giving E=mc² / √(1 – $v^2/c^2$). Einstein's 1905 "inertia" paper [10] gives a more general calculation for an opposing pair of plane waves tilted at any angle in the emitter-frame.

Since the angle-dependent information needed to calculate the momentum components of light-rays in the first two elliptical maps differs only by a Lorentz scaling, the angle-independence of the 1905 derivation also applies to our "emitter-theory" derivation.

### Reworking special relativity

Ellipse 3(ii) contains the same key relationships as special relativity, but without explicitly using the special theory's assumption of flat spacetime, or by dividing the phenomena into separate "propagation" and "Lorentz" components.

It is possible that other researchers may be interested in examining a class of model in which local lightspeed constancy is regulated by spacetime distortions associated with the relative motion of physical particles (e.g. [11]), but which generates the same basic shift predictions as special theory.

We do not expect to see a theory based on 3(iii).

## 9. Conclusions

The wavefront maps presented in section 3 can be useful as a visualisation aid for relativistic problems.

Investigating the properties of these maps raised three additional points:

- Special relativity's aberration formula can be calculated from general principles without committing to special relativity's shift equations.

- Transverse redshifts can also be derived from a first-order Doppler equation, although the strength of the effect is different under special relativity.

- Special relativity's key relationships can be deduced from the principle of relativity and the "relativistic Doppler" shift equation without assuming that spacetime is flat, suggesting the possibility of a "non-flat" variation on special relativity.

Although these points are probably not new, they can be difficult to derive by other means.